\input psfig.sty

\def\ptitle{Energy bounds for the spinless Salpeter equation}
\def\stitle{harmonic oscillator}
\nopagenumbers
\magnification=\magstep1
\hsize 6.0 true in 
\hoffset 0.25 true in 
\emergencystretch=0.6 in 
\vfuzz 0.4 in 
\hfuzz 0.4 in 
\vglue 0.1true in
\mathsurround=2pt 
\topskip=20pt 
\def\nl{\noindent} 
\def\nll{\hfil\break\noindent} 
\def\np{\hfil\vfil\break} 
\def\title#1{\bigskip\noindent\bf #1 ~ \trr\smallskip} 
\def\htab#1#2{{\hskip #1 in #2}}
\font\trr=cmr10 
\font\bf=cmbx10 
\font\bmf=cmmib10 
\font\sl=cmsl10 
\font\it=cmti10 
\font\trbig=cmbx10 scaled 1500 
\font\tiny=cmr8 
\def\mb#1{\hbox{\bmf#1}} 
\def\ng{>\kern -9pt|\kern 9pt} 
\def\hi#1#2{$#1$\kern -2pt-#2} 
\def\hy#1#2{#1-\kern -2pt$#2$} 

\def\sgn{{\rm sgn}}
\def\half{{1 \over 2}}


\output={\shipout\vbox{\makeheadline
\ifnum\the\pageno>1 {\hrule} \fi 
{\pagebody} 
\makefootline}
\advancepageno}

\headline{\noindent {\ifnum\the\pageno>1 
{\tiny \ptitle\hfil page~\the\pageno}\fi}}
\footline{}
\newcount\zz \zz=0 
\newcount\q 
\newcount\qq \qq=0 

\def\pref #1#2#3#4#5{\frenchspacing \global \advance \q by 1 
\edef#1{\the\q}
{\ifnum \zz=1 { %
\item{[\the\q]} 
{#2} {\bf #3},{ #4.}{~#5}\medskip} \fi}}

\def\bref #1#2#3#4#5{\frenchspacing \global \advance \q by 1 
\edef#1{\the\q}
{\ifnum \zz=1 { %
\item{[\the\q]} 
{#2}, {\it #3} {(#4).}{~#5}\medskip} \fi}}

\def\gref #1#2{\frenchspacing \global \advance \q by 1 
\edef#1{\the\q}
{\ifnum \zz=1 { %
\item{[\the\q]} 
{#2}\medskip} \fi}}

\def\sref #1{~[#1]}

\def\references#1{\zz=#1
\parskip=2pt plus 1pt 
{\ifnum \zz=1 {\noindent \bf References \medskip} \fi} \q=\qq
\pref{\bse}{E.~E.~Salpeter and H.~A.~Bethe, Phys.~Rev.}{84}{1232 (1951)}{}

\pref{\se}{E.~E.~Salpeter, Phys.~Rev.}{87}{328 (1952)}{}
\pref{\luca}{W.~Lucha and F.~F.~Sch\"oberl, Phys.~Rev.\ A}{54}{3790 (1996)}{}
\pref{\lucb}{W.~Lucha and F.~F.~Sch\"oberl, Int.~J.~Mod.~Phys.\ A}{14}{2309
(1999)}{}
\pref{\lucc}{W.~Lucha and F.~F.~Sch\"oberl, Fizika B}{8}{193 (1999)}{}
\pref{\lucd}{W.~Lucha and F.~F.~Sch\"oberl, Phys.~Rev.\ A}{60}{5091 (1999)}{}
\pref{\luce}{W.~Lucha and F.~F.~Sch\"oberl, Int.~J.~Mod.~Phys.\ A}{15}{3221
(2000)}{}
\bref{\reed}{M. Reed and B. Simon}{Methods of Modern Mathematical Physics IV: Analysis of Operators}{Academic, New York, 1978}{}
\pref{\halla}{R. L. Hall, J. Math. Phys.}{24}{324 (1983)}{} 
\pref{\hallb}{R. L. Hall, J. Math. Phys.}{25}{2078 (1984)}{} 
\pref{\hallc}{R. L. Hall, J. Math. Phys.}{34}{2779 (1993)}{} 
\pref{\hallw}{R. L. Hall, J. Phys. G}{26}{981 (2000)}{} 

}

\references{0} 

\topskip=0pt 

\trr 
\htab{3}{HEPHY-PUB 735/00}

\htab{3}{UWThPh-2000-51}

\htab{3}{CUQM-82}

\htab{3}{hep-th/0012127}

\htab{3}{December 2000}
\vskip 0.8 true in
\centerline{\trbig \ptitle:}
\vskip 0.2true in
\centerline{\trbig \stitle}
\vskip 0.5true in
\baselineskip 12 true pt 
\centerline{\bf Richard L. Hall$^{1}$, Wolfgang Lucha$^{2}$, and Franz F. Sch\"oberl$^{3}$}\medskip
\nll $^{(1)}${\sl Department of Mathematics and Statistics, Concordia
University, 1455 de Maisonneuve Boulevard West, Montr\'eal, Qu\'ebec, Canada H3G 1M8}
\nll $^{(2)}${\sl Institut f\"ur Hochenergiephysik, \"Osterreichische Akademie der Wissenschaften, Nikolsdorfergasse 18, A-1050 Wien, Austria}
\nll $^{(3)}${\sl Institut f\"ur Theoretische Physik, Universit\"at Wien, Boltzmanngasse 5, A-1090 Wien, Austria}

\nll{email: \sl rhall@cicma.concordia.ca, wolfgang.lucha@oeaw.ac.at, franz.schoeberl@univie.ac.at}
\bigskip\bigskip

\baselineskip = 18true pt 
\topskip=20pt 
\centerline{\bf Abstract}\medskip
\noindent We study the eigenvalues ${\cal E}_{n\ell}$ of the Salpeter
Hamiltonian $H = \beta\sqrt{m^2 + \mb{p}^2} + vr^2,$ $v>0,$ $\beta>0,$ in
three dimensions. By using geometrical arguments we show that, for suitable
values of $P,$ here provided, the simple semi-classical formula $${\cal
E}_{n\ell} \approx \min_{r > 0} \left\{v(P_{n\ell}/r)^2 + \beta\sqrt{m^2 +
r^2}\right\}$$
\nl provides both upper and lower energy bounds for all the eigenvalues of
the problem.

\np
\title{1.~~Introduction}
\noindent The Bethe-Salpeter formalism\sref{\bse} is generally accepted, in
principle, as the appropriate framework for the description of bound states
within a (relativistic) quantum field theory. Unfortunately, almost all
applications of this formalism face serious problems of both conceptual and
practical nature. In particular, it turns out to be a highly nontrivial task
to extract exact information about the solutions. Consequently, one is led
to consider some reasonable simplifications of the full Bethe-Salpeter
equation, such as the following: the elimination of any dependence on
timelike variables by the use of a static or instantaneous approximation to
the interaction kernel, leading to the `Salpeter equation'\sref{\se}; the
neglect of the spin degrees of freedom of the bound-state constituents; and
the restriction to positive-energy solutions. With these constraints one
arrives at the `spinless Salpeter equation', which may be regarded as a
straightforward generalization of the nonrelativistic Schr\"odinger formalism
towards relativistic kinematics: it describes the bound states of scalar
particles as well as the spin-averaged spectra of the bound states of
fermions.

In this paper we study the Salpeter Hamiltonian $$H = \beta\sqrt{m^2 +
\mb{p}^2} + v r^2\eqno{(1.1)}$$
\nl in which $\beta > 0$ is a parameter (allowing, for example, for more than one particle), $m$ is the mass, and $vr^2$ is the
harmonic-oscillator potential with coupling $v > 0.$ In the momentum-space
representation\sref{\luca-\luce} the operator $\mb{ p}$ becomes a
\hi{c}{variable} and thus, from the spectral point of view, the Hamiltonian
$H = \beta\sqrt{m^2 + \mb{p}^2} + v r^2$ is equivalent to the Schr\"odinger
operator ${\cal H}$ given by $${\cal H} = -v\Delta + V(r),\quad V(r) =
\beta\sqrt{m^2 + r^2}.\eqno{(1.2)}$$
\nl Since the potential $V(r)$ in (1.2) increases without bound, we
know\sref{\reed} that the spectrum of the operator ${\cal H}$ is entirely
discrete, and we denote its eigenvalues (also the eigenvalues of $H$) by
$${\cal E}_{n \ell}(v, \beta, m),\quad n = 1,2,3,\dots,\ \ell =
0,1,2,\dots,\eqno{(1.3)}$$
\nl where $n$ `counts' the radial states in each angular-momentum subspace,
labelled by $\ell.$ Because $V(r)$ is at once a concave function of $r^2$ and
a convex function of $r,$ this allows us to derive in Sec.~2 the
approximation formula $${\cal E}_{n\ell} \approx \min_{r > 0}
\left\{v\left({{P_{n\ell}}\over{r}}\right)^2 + \beta\sqrt{m^2 + r^2}
\right\},\eqno{(1.4)}$$ which, for suitable values of $P,$ provides both
upper and lower bounds to the energy.
\title{2.~~The energy bounds}
\noindent If we think of the potential $V(r)$ as a smooth transformation $V(r) =
g(\sgn(q)r^q)$ of a pure attractive power potential $\sgn(q)r^q,$ then the
potential $V(r)$ clearly has two such representations, each with definite
convexity. That is to say, we may write the potential in the following way:
$$V(r) = \beta\sqrt{m^2 + r^2} = g^{(1)}(r^2) = g^{(2)}(r),\eqno{(2.1)}$$
\nl where the two transformation functions $g$ have the properties that $g^{(1)}$
is concave ($g'' < 0$) and $g^{(2)} = V$ is convex ($g'' > 0$). These are
precisely the conditions under which our `envelope
theory'\sref{\halla-\hallc} applies. The tangent lines to the transformation
functions $g$ are, on the one hand, shifted oscillator potentials of the form
$a + br^2,$ and, on the other, shifted linear potentials of the form $a + br.$
The potential $V(r)$ itself is at once the envelope of an upper oscillator
family and a lower linear family of potentials. It follows\sref{\hallc} that
we may write $$\min_{r > 0} \left\{v\left({{P_{n\ell}}(1)\over{r}}\right)^2 +
\beta\sqrt{m^2 + r^2} \right\} \leq {\cal E}_{n\ell}\leq \min_{r > 0}
\left\{v\left({{P_{n\ell}}(2)\over{r}}\right)^2 + \beta\sqrt{m^2 +
r^2}\right\},
\eqno{(2.2)}$$ 
\nl where the `upper' and `lower' $P$ numbers are defined in terms of the
corresponding exact power eigenvalues $E$ as follows. Suppose the exact
eigenvalues of the Schr\"odinger operator for the pure-power potential
$-\Delta + \sgn(q)r^q,$ $q \geq -1,$ are written $E_{n\ell}(q),$ then the
corresponding $P$ numbers are defined\sref{\hallc} by $$P_{n\ell}(q) =
\left|E_{n\ell}(q)\right|^{(2+q)/2q}\left[{2\over{2+q}}\right]^{1/q}\left[{{|q|}\over{2+q}}\right]^{1/2},\quad
q \neq 0.\eqno{(2.3)}$$
\nl The limiting case $q\rightarrow 0$ corresponds {\it exactly} to the
$\log(r)$ potential, but that is another story\sref{\hallw}. Thus we have for
the harmonic-oscillator potential ($q = 2$) $$E_{n\ell}(2) = 4n + 2\ell -
1,\quad\quad P_{n\ell}(2) = 2n + \ell - \half,\eqno{(2.4)}$$
\nl and for the linear potential ($q=1$) $$P_{n\ell}(1) =
2\left({{E_{n\ell}(1)}\over{3}}\right)^{3/2},\eqno{(2.5)}$$
\nl which are listed here in Table~1. Thus we have established the
principal claim of this paper. The minimizations in (2.2) can be carried out
exactly, yielding the following quartic equation for $r^2$:$$r^8 =
{{4v^2P^4}\over{\beta ^2}}(m^2 + r^2).\eqno{(2.6)}$$
\nl However, given the ease of contemporary computing, the minimizations in
(2.2) are perhaps preferable to the exact analytical solutions to (2.6);
moreover, (2.2) involves meaningful semi-classical energy expressions which
exhibit clearly the kinetic and potential energy contributions and how they
depend on the parameters of the problem. It turns out that these energy
bounds are quantitatively equivalent to the bounds obtained in the Appendix
of Reference\sref{\luce} by the use of optimized operator inequalities. The
present formulation of the energy bounds, based on convexity, allows for the
uniform and succinct expression of our upper and lower results by Eq.~(2.2),
and admits further natural generalizations.

As an illustration of the accuracy of (2.2) we have plotted graphs of the
approximate eigenvalues given by (2.2) in Fig.~1. We illustrate the case
$\beta = v = 1$ and plot the eigenvalue bounds obtained as functions of the
mass $m.$ Corresponding accurate numerical eigenvalues are shown as a dashed
curve between each pair of bound curves. These graphs confirm numerically
what is immediately clear directly from the operator $H,$ that the
corresponding (Schr\"odinger) problem ${\cal H}$ approaches the oscillator
for large $m,$ and the linear potential for small $m.$

\title{3.~~Conclusion}
\noindent The spinless-Salpeter eigenvalue problem is not easy to solve. Even for the
harmonic-oscillator potential, one obtains an equivalent Schr\"odinger
problem with the potential $V(r) = \beta\sqrt{m^2 + r^2}$ that does not admit
exact analytical solutions in terms of known special functions. Thus, even
for the oscillator problem, we must resort to approximations of some sort.
The potential $V(r)$ increases monotonically to infinity and we therefore
know {\it a priori} that the spectrum is entirely discrete. Hence, the
spectrum and wave functions can be found numerically with considerable ease.
In spite of this, it is always desirable to have an eigenvalue {\it formula},
even an approximate one, which tells us how the spectrum depends on all the
parameters of the problem. In this paper we have used geometrical envelope
theory to generate simple semi-classical expressions that provide upper and
lower bounds to all the eigenvalues.

\title{Acknowledgment}
Partial financial support of this work under Grant No. GP3438 from the Natural Sciences and Engineering Research Council of Canada, and hospitality of the Erwin Schr\"odinger International Institute for Mathematical Physics in Vienna is gratefully acknowledged by one of us [RLH]. 
\np
\references{1}
\np
\noindent {\bf Table 1}~~Numerical values for the $P$ numbers for the linear potential ($q = 1$) used in the Schr\"odinger eigenvalue formula (2.3) 
\baselineskip=16 true pt 
\def\vr{\vrule height 12 true pt depth 6 true pt}
\def\vra{\vr\hfill} \def\vrb{\hfill &\vra} \def\vrc{\hfill & \vr\cr\hrule}
\def\vrq{\vr\quad} 

$$\vbox{\offinterlineskip
\hrule
\settabs
\+ \vrq \kern 0.4true in &\vrq \kern 0.4true in &\vrq \kern 0.7true in &\vrq \kern 0.7true in &\vrq \kern 0.5true in &\vr\cr\hrule
\+ \vra $n$ \vrb $\ell$\vrb $P_{n\ell}(1)$\vrc
\+ \vra 1\vrb 0\vrb 1.37608\vrc 
\+ \vra 2\vrb 0\vrb 3.18131\vrc 
\+ \vra 3\vrb 0\vrb 4.99255\vrc
\+ \vra 4\vrb 0\vrb 6.80514\vrc
\+ \vra 5\vrb 0\vrb 8.61823\vrc
\+ \vra 1\vrb 1\vrb 2.37192\vrc
\+ \vra 2\vrb 1\vrb 4.15501\vrc
\+ \vra 3\vrb 1\vrb 5.95300\vrc
\+ \vra 4\vrb 1\vrb 7.75701\vrc
\+ \vra 5\vrb 1\vrb 9.56408\vrc
\+ \vra 1\vrb 2\vrb 3.37018\vrc
\+ \vra 2\vrb 2\vrb 5.14135\vrc
\+ \vra 3\vrb 2\vrb 6.92911\vrc
\+ \vra 4\vrb 2\vrb 8.72515\vrc
\+ \vra 5\vrb 2\vrb 10.52596\vrc
\+ \vra 1\vrb 3\vrb 4.36923\vrc
\+ \vra 2\vrb 3\vrb 6.13298\vrc
\+ \vra 3\vrb 3\vrb 7.91304\vrc
\+ \vra 4\vrb 3\vrb 9.70236\vrc
\+ \vra 5\vrb 3\vrb 11.49748\vrc
\+ \vra 1\vrb 4\vrb 5.36863\vrc
\+ \vra 2\vrb 4\vrb 7.12732\vrc
\+ \vra 3\vrb 4\vrb 8.90148\vrc
\+ \vra 4\vrb 4\vrb 10.68521\vrc
\+ \vra 5\vrb 4\vrb 12.47532\vrc
}$$
\np
\baselineskip 18 true pt 


\hbox{\vbox{\psfig{figure=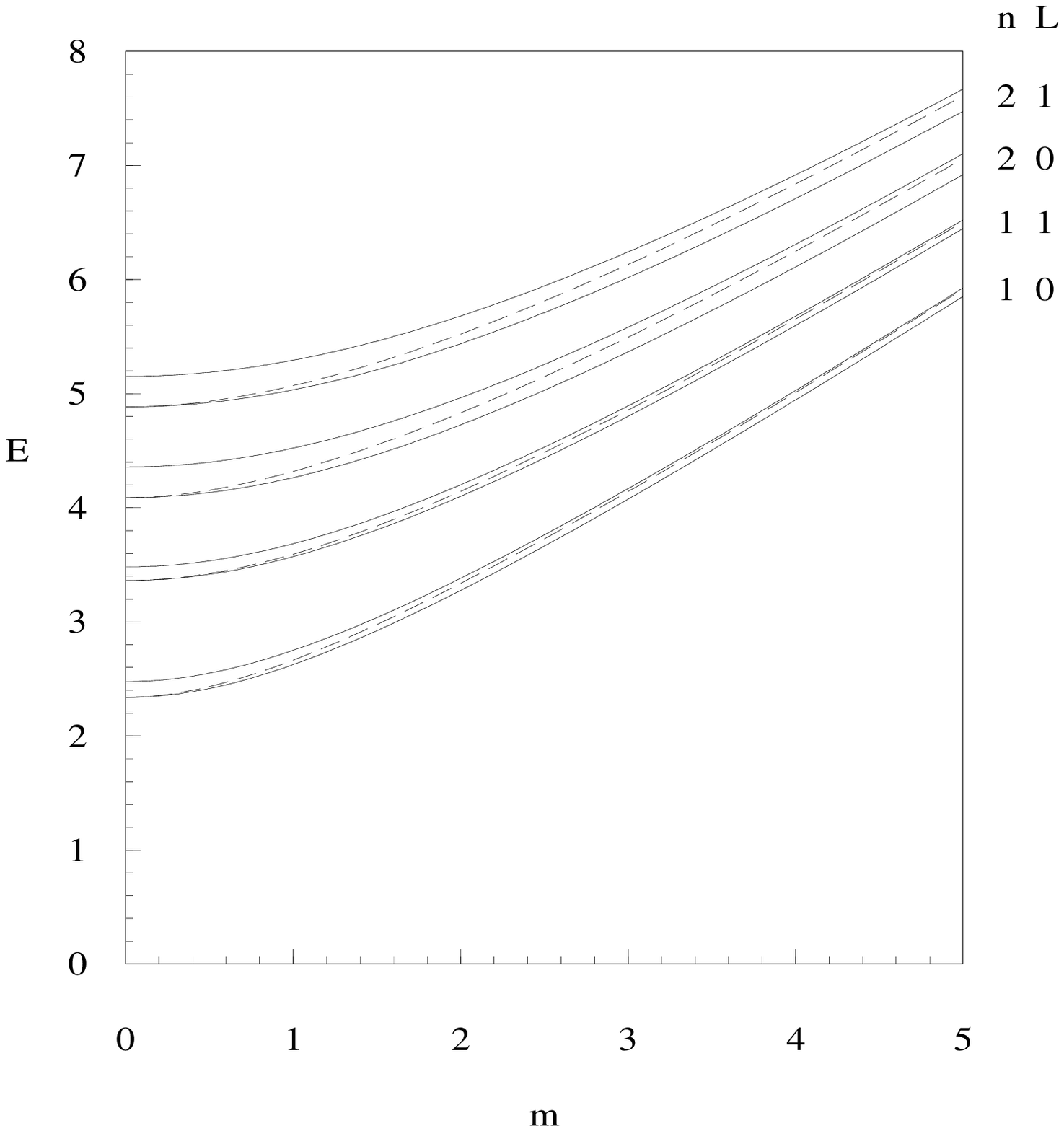,height=6in,width=5in,silent=}}}

\title{Figure 1.}
\nl Upper and lower bounds on the energy levels of the spinless Salpeter
Hamiltonian $H$ with a harmonic-oscillator potential,
$H=\beta\sqrt{m^2+\mb{p}^2}+vr^2,$ for $\beta=v=1,$ as functions of the mass
$m.$ Accurate numerical eigenvalues are shown as dashed curves between each
pair of bounds.


\end